\documentclass[aps,prb,twocolumn,superscriptaddress,showpacs]{revtex4-1}
\usepackage{graphicx}
\begin{document}

\title{Transport Measurements of Strongly-Correlated Electrons on Helium in a Classical Point-Contact Device}


\author{D.G. Rees}
\email[]{drees@riken.jp}
\affiliation{Low Temperature Physics
Laboratory, RIKEN, Wako 351-0198, Japan}

\author{I. Kuroda}
\affiliation{Low Temperature Physics
Laboratory, RIKEN, Wako 351-0198, Japan}
\affiliation{Department of Physics, Tokyo Institute of Technology, Tokyo 152-8551, Japan}

\author{C.A. Marrache-Kikuchi}
\altaffiliation{Present address: CSNSM, Bat. 108, 91405 Orsay, France}
\affiliation{Low Temperature Physics
Laboratory, RIKEN, Wako 351-0198, Japan}

\author{M. H\"{o}fer}
\affiliation{Faculty of Physics, University of Konstanz, 78457 Konstanz, Germany}

\author{P. Leiderer}
\affiliation{Faculty of Physics, University of Konstanz, 78457 Konstanz, Germany}

\author{K. Kono}
\affiliation{Low Temperature Physics
Laboratory, RIKEN, Wako 351-0198, Japan}
\affiliation{Department of Physics, Tokyo Institute of Technology, Tokyo 152-8551, Japan}



\date{\today}
\begin{abstract}
We present transport measurements of electrons on the
surface of liquid helium in a microchannel device in which a
constriction may be formed by a split-gate
electrode. The surface electron current passing
through the microchannel first decreases and is then completely
suppressed as the split-gate voltage is swept negative. The current decreases in a steplike manner, due to changes in the number of electrons able to pass simultaneously through the constriction. We investigate the dependence of the electron transport on the AC driving voltage and the DC potentials applied to the
sample electrodes, in order to understand the
electrostatic potential profile of the constriction region. Our results are in good agreement with a finite element modeling analysis of the device. We demonstrate that the threshold of current flow depends
not only on the applied potentials but also on the surface electron
density. The detailed understanding of the characteristics of such
a device is an important step in the development of mesoscopic
experiments with surface electrons on liquid helium.
\end{abstract}
\maketitle
\section{Introduction}
\label{intro} An electron close to the free surface of liquid helium
is subjected to an attractive force due to a weak image charge
formed in the liquid\cite{Andrei,MonarkhaKono}. In the presence of an applied electric field
$E_z$, the potential in the direction perpendicular to the helium
surface may be written as $V(z) = -\Lambda e^2/z + eE_zz$ where
$\Lambda = (\varepsilon - 1)/4(\varepsilon + 1)$, $\varepsilon$
is the dielectric constant of the liquid, $e$ is the absolute value of the electronic charge and $z$ is the distance from the helium surface. The electron is
prevented from entering the liquid due to a $\sim 1$ eV potential
barrier at the surface and so remains localized above the liquid. The potential $V(z)$ gives rise to a series of bound
states for perpendicular motion\cite{GrimesImageStates}; in the ground state, the
expectation value of $z$ is $\sim 11$ nm. The lifetime of these
bound states is predicted to be long, leading to the proposal that
electrons on the surface of liquid helium may be good candidates
for quantum bits\cite{PlatzmanandDykmanscience}.

For a two-dimensional system of charges on a liquid helium
surface, the electron surface density $n_s$ may be varied over a
wide range, up to a theoretical limit given by the hydrodynamic
instability of the bulk liquid surface\cite{Gorkov,Leiderer1979189,LeidererDimples} of $\sim$2.2$\times10^9$
cm$^{-2}$. By varying temperature the scattering processes which
determine the electron mobility in the plane parallel to the
helium surface may be controlled. Below $\sim0.8$ K the density of
gas atoms above the liquid surface becomes effectively zero and
scattering occurs only with excitations of the liquid surface,
ripplons, leading to high mobilities\cite{ShirahamaMobility} in excess of 10$^8$
cm$^2/$V$\cdot$s. As the Coulomb interaction between electrons is
essentially unscreened, and the electron separation is much larger
than the thermal electron wavelength, SSE have been used to study
classical effects in strongly-interacting electron systems, such as
the transition from an electron liquid to a 2D Wigner solid as the
temperature of the electron system is decreased\cite{GrimesAdamsWignerCrystal}.

Quasi-one dimensional SSE systems\cite{KovdryaReview} have been investigated using
samples in which liquid helium was confined in grooves on a
dielectric substrate\cite{KovdryaNikolaenko}. The width of the grooves was as low as 1.25
$\mu$m. In such experiments, the temperature dependence of the electron mobility\cite{KovdryaNikolaenko2} was
found to be in good agreement with that predicted by theoretical
calculations where transitions between quantized energy
levels for the lateral motion of the electrons were taken into
account\cite{Sokolov1Dmobility}. However, for such dielectric substrates the mobility was
also dependent on substrate defects in regions where the helium film
was thin.

\begin{figure}
  \includegraphics[angle=0,width=0.45\textwidth]{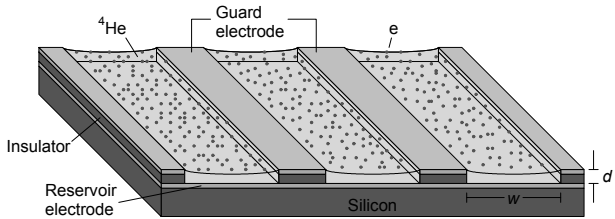}
\caption{Schematic picture of a set of microchannels. The microchannels are filled by the capillary action of
superfluid $^4$He, the surface of which is then charged with electrons.}
\label{fig:1}
\end{figure}

Recent experiments have made use of microchannel devices
fabricated by lithographic techniques in order to study the
properties of SSE in confined geometries. A schematic
representation of a set of such microchannels is shown in Fig. \ref{fig:1}. Placed a distance $h$ above the bulk surface of superfluid helium,
the channel will fill by capillary action; the helium surface may
then be charged. The radius of curvature $R$ of the liquid is
given by $R = \alpha/\rho g h +
n_s^2e^2/2\varepsilon\varepsilon_0$ where $\alpha$ and $\rho$ are
the surface tension coefficient and density of liquid $^4$He
respectively, $g$ is the acceleration due to gravity and $\varepsilon_0$ is the dielectric constant of vacuum\cite{MartyCurvature}. The
transport of electrons in such devices was first demonstrated for
a channel of width 30 $\mu$m and depth 1 $\mu$m\cite{vanHaren}. The non-linear
transport of the Wigner solid on helium surfaces in microchannels
of width 8-20 $\mu$m was studied in three-terminal devices
comprising of source, drain and gate electrodes submerged beneath
the helium surface\cite{Phil'sChannels,IkegamiWigner}. The ultra-efficient transfer of a small number
of electrons along parallel microchannels of width 10 $\mu$m,
again using a series of gate electrodes beneath the helium
surface, has also been demonstrated\cite{sabouret}. In still more advanced
devices, small ensembles of electrons, including a single
electron, trapped in a microfabricated circular pool of radius 10
$\mu$m, have been studied using a charge-sensitive superconducting
Single Electron Transistor (SET) positioned beneath the helium
surface\cite{CountingElectronsOnLHe}. A Field Effect Transistor (FET) for electrons on a thin
helium film, where the electron density may exceed the
hydrodynamic limit, has also been demonstrated\cite{heliumFET}. There the
separation between the split-gate electrodes was 200 $\mu$m.

Such experiments demonstrate a progression towards the sensitive
control and measurement of small numbers of SSE on helium, raising
the possibility of studying novel phenomena associated with
classically interacting charge systems in confined geometries.
Many theoretical studies have been conducted on the behavior of
such systems. In quasi-one dimensional systems, at sufficiently
low temperatures, particles are predicted to form a series of
rows, the number of which changes with the particle density or
confinement strength, leading to structural phase transitions and
re-entrant melting processes\cite{Chaplik,Peeters1DCrystal}. Similar phenomena have also been
predicted in circularly symmetric parabolic confinements\cite{bedanov}. The
pinning and depinning dynamics of charged particles at potential
constrictions have also been investigated using Monte Carlo
calculations\cite{PeetersConstriction,daSilvaPinning,damascenoPinning}. However, experimental difficulties have restricted
progress towards investigating these phenomena. SET charge
measurements, whilst extremely sensitive, may be plagued by
intrinsic two-level fluctuator charge noise\cite{Zimmerman,reesSCPT}. Also, the
electrostatic potential profile in microchannel devices may be
distorted by contact potentials and surface charging effects which
can be difficult to quantify\cite{mukharsky}.

As a step towards overcoming these experimental difficulties, we
have performed transport measurements of SSE on superfluid $^4$He
in a microfabricated device. In this sample, two microchannel
SSE reservoirs are separated by a split-gate electrode which, at
appropriate bias, forms a constriction. As the split-gate
voltage $V_{gt}$ is swept negative the current \emph{I} flowing
through the constriction is reduced and then reaches zero at a
threshold voltage $V^{th}_{gt}$. The appearance of step-like decreases in current in this device, where each step corresponds to a change in the number of electrons able to pass simultaneously through the constriction, has already been reported\cite{Reesprl}. Here we present further transport measurements, investigating in particular the response of the system to increasing AC driving voltage, which causes significant non-linear transport effects to emerge. These results are discussed in relation to an electrostatic model of the sample which shows that offsets in electrode potentials, arising for reasons which are not yet clear, are important in determining the potential profile of the device.

\begin{figure}
  \includegraphics[angle=-90,width=0.5\textwidth]{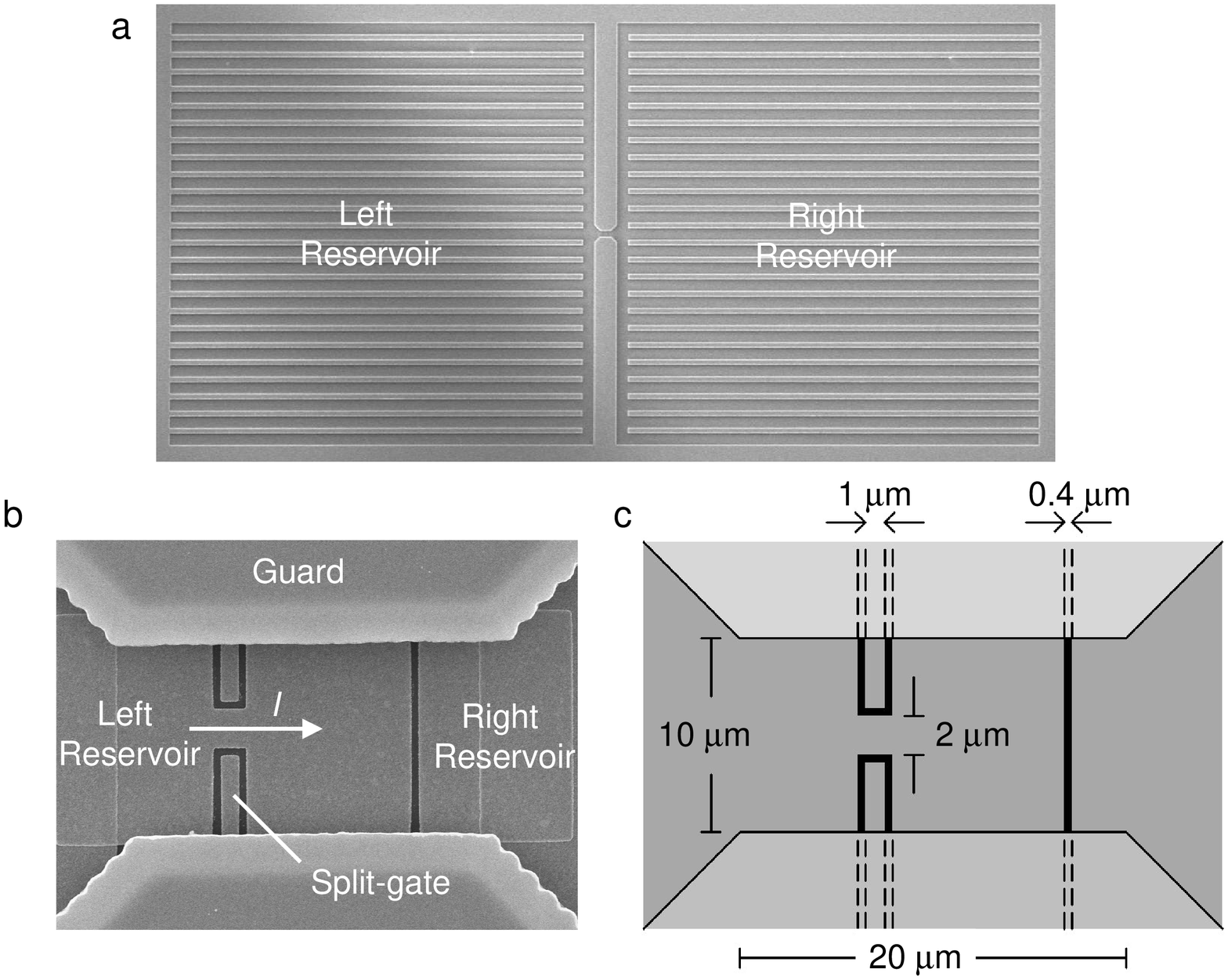}
\caption{Point-contact device for electrons on helium. (a) Two arrays of microchannels form
the left and right electron reservoirs. (b) A split-gate electrode is positioned at the base of the small central channel which links the two reservoirs. (c) Diagram showing the dimensions of the electrodes in the central channel.}
\label{fig:2}
\end{figure}

\section{Experimental}
\label{Exp} The sample used in this experiment was prepared using
multilayer optical and electron-beam lithography on a Si wafer,
the surface of which was oxidized. Two arrays of microchannels,
defined by a guard electrode, act as electron reservoirs between
which electrons may be exchanged. These left and right reservoirs
consist of 25 microchannels of width $w =$ 20 $\mu$m arranged in
parallel and connected together at one end (Fig. \ref{fig:2}(a)). The two reservoirs are
separated by a smaller channel of width 10 $\mu$m and length 20
$\mu$m. Electrodes were fabricated beneath the reservoir
microchannels and are denoted as the left and right reservoir
electrodes respectively. A split-gate electrode was fabricated at
the base of the small central channel (Fig. \ref{fig:2}(b)). The geometry of
the central channel is shown in Fig. \ref{fig:2}(c); the split gate was 1
$\mu$m long and separated by a gap of 2.8 $\mu$m. The 400 nm gap
between left and right reservoir electrodes was placed 10 $\mu$m
to the right of the split gate in order to avoid the
distortion of the potential profile of the constriction region. A 1.5 $\mu$m thick
layer of hard baked photoresist, which defined the microchannel
depth $d$, separated the lower electrodes from the guard
electrode. All metal layers were made of gold (65 nm thick) on top
of a thin (15 nm) titanium layer which was deposited in order to promote adhesion.

The sample was placed in an experimental cell approximately 0.5 mm
above the bulk surface of superfluid $^4$He at 1.25 K. The potentials $V_{gu}$,
$V_{r}$ and $V_{gt}$ were applied to the guard, reservoir and split-gate electrodes, respectively. Under the bias conditions
$V_{gu}$ = 0 V, $V_{r} = V_{gt} = +1.0$ V, the surface of the
helium was charged by thermionic emission from a small tungsten
filament placed a few mm above the sample. A small AC voltage
$V_{in}$ of frequency 200 kHz was superimposed on the right
reservoir electrode in order to drive electrons between the two reservoirs,
through the central channel. The current \emph{I} and conductance
of the electron system \emph{G} were determined by making a
phase-sensitive measurement of the voltage capacitively induced on
the left reservoir electrode, in reference to the standard
lumped-circuit model\cite{Iye}.

\begin{figure}
  \includegraphics[angle=0,width=0.25\textwidth]{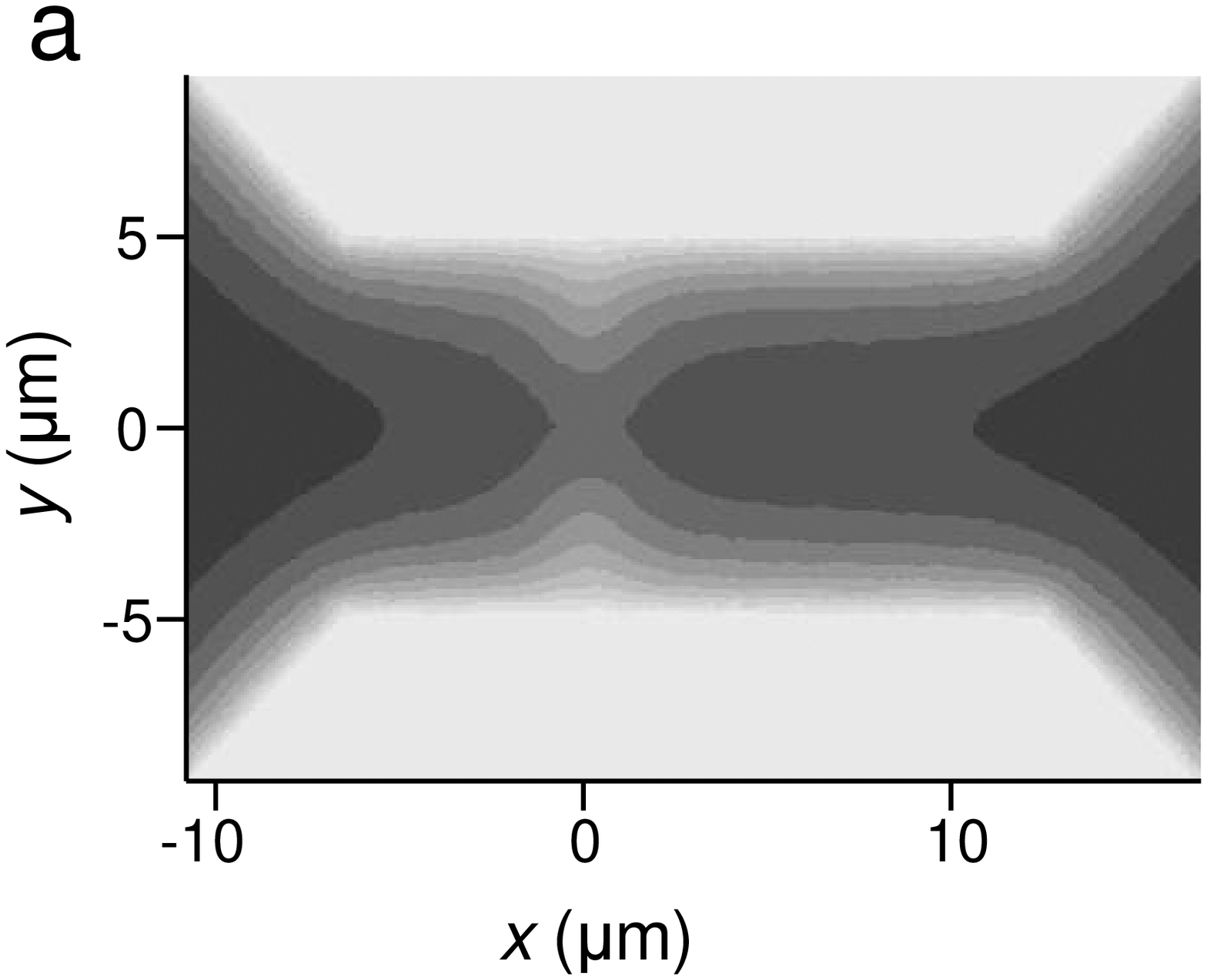}
  \includegraphics[angle=0,width=0.225\textwidth]{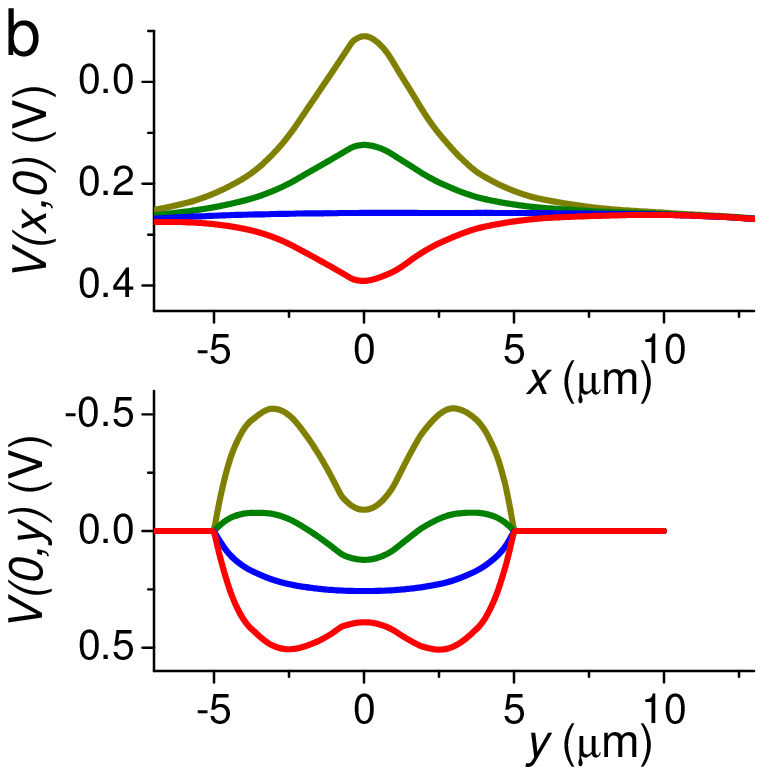}
\caption{(Color online) Finite element modeling results. (a) Contour plot of the calculated electrostatic potential in the central channel for $V_{gu}=V_{gt}=0$ V and $V_{r}=+1.0$ V. The darker areas correspond to regions of lower potential energy for electrons. (b) Electrostatic potential in the $x$ direction along the channel ($y=0$ $\mu$m) and the $y$ direction across the channel ($x=0$ $\mu$m), for $V_{gu}=0$ V, $V_{r}=0.3$ V and $V_{gt}=$+1.3 V (red), +0.3 V (blue), -0.7 V (green), -2.3 V (dark yellow).}
\label{fig:3}
\end{figure}

To aid the understanding of the electrostatic potential profile of
the device, a finite element model of the central channel was
developed\cite{Flex}. The results of the modeling are
shown in Fig. \ref{fig:3}(a) for $V_{gu} = V_{gt} = 0$ V, $V_{r} = +0.3$ V.
When $V_{gt}$ is more negative than $V_r$, a saddle-point potential is formed on the helium
surface at the point ($x=0,y=0$), with a maximum in potential for electrons in the
\emph{x} direction along the channel and a minimum in the lateral
\emph{y} direction. Fig. \ref{fig:3}(b) shows how the potential profile
develops as $V_{gt}$ is changed. For $V_{gt}=$-2.3 V the potential at the center of the saddle-point is more negative than the guard electrode potential. We assume that the electrostatic potential energy of the electron system may not exceed $V_{gu}$, as this would lead to electrons escaping from the reservoirs, onto the thin helium film covering the guard electrode, through which they should rapidly drain away\cite{LeidererThinFilm}. Therefore, when the potential at the center of the saddle-point is more negative than the guard electrode potential, the resulting potential barrier between the two reservoirs should block electron transport through the constriction.

\section{Results and Discussion}
\label{Res and dis}
\begin{figure}
  \includegraphics[angle=0,width=0.23\textwidth]{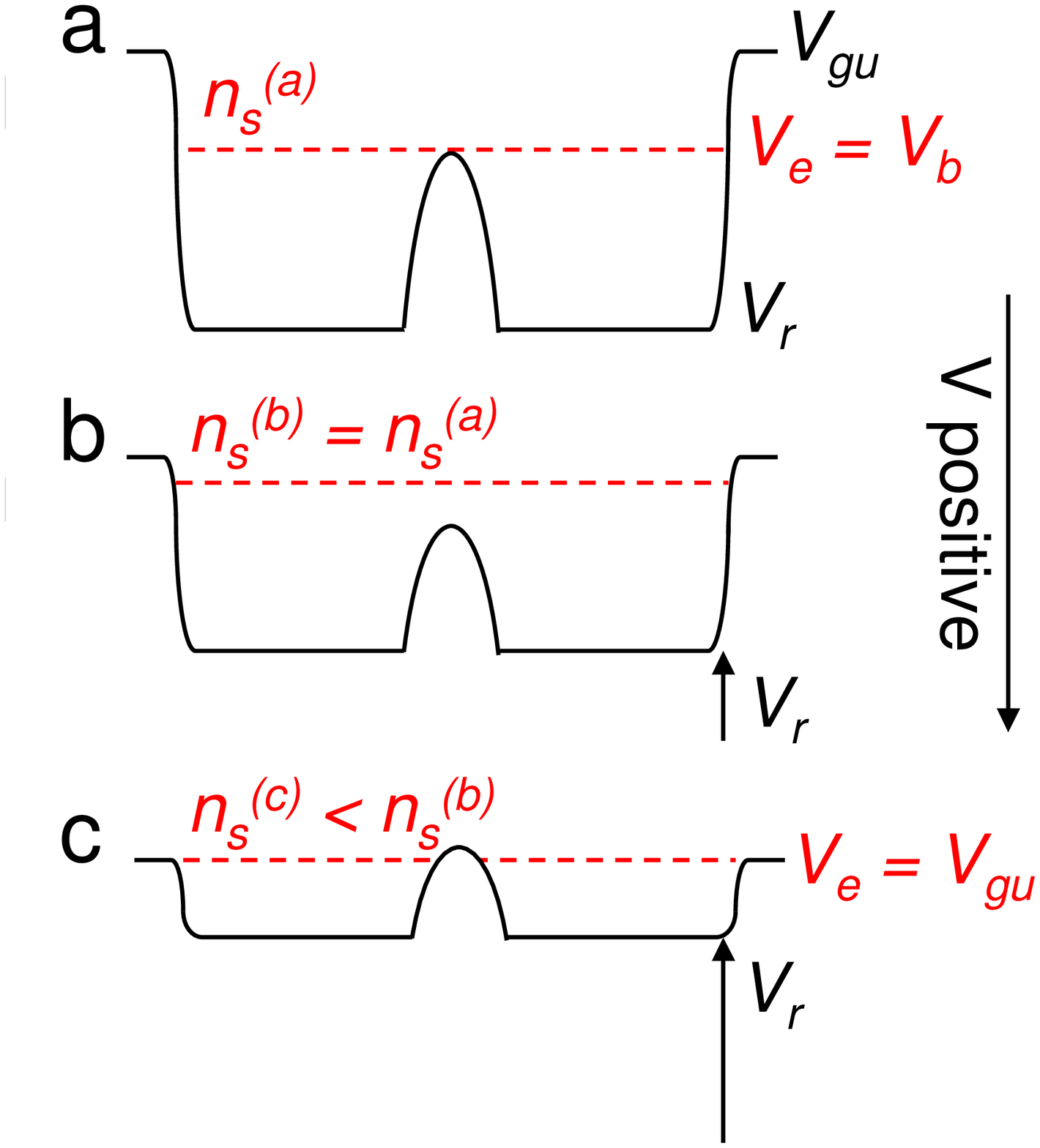}
    \includegraphics[angle=0,width=0.23\textwidth]{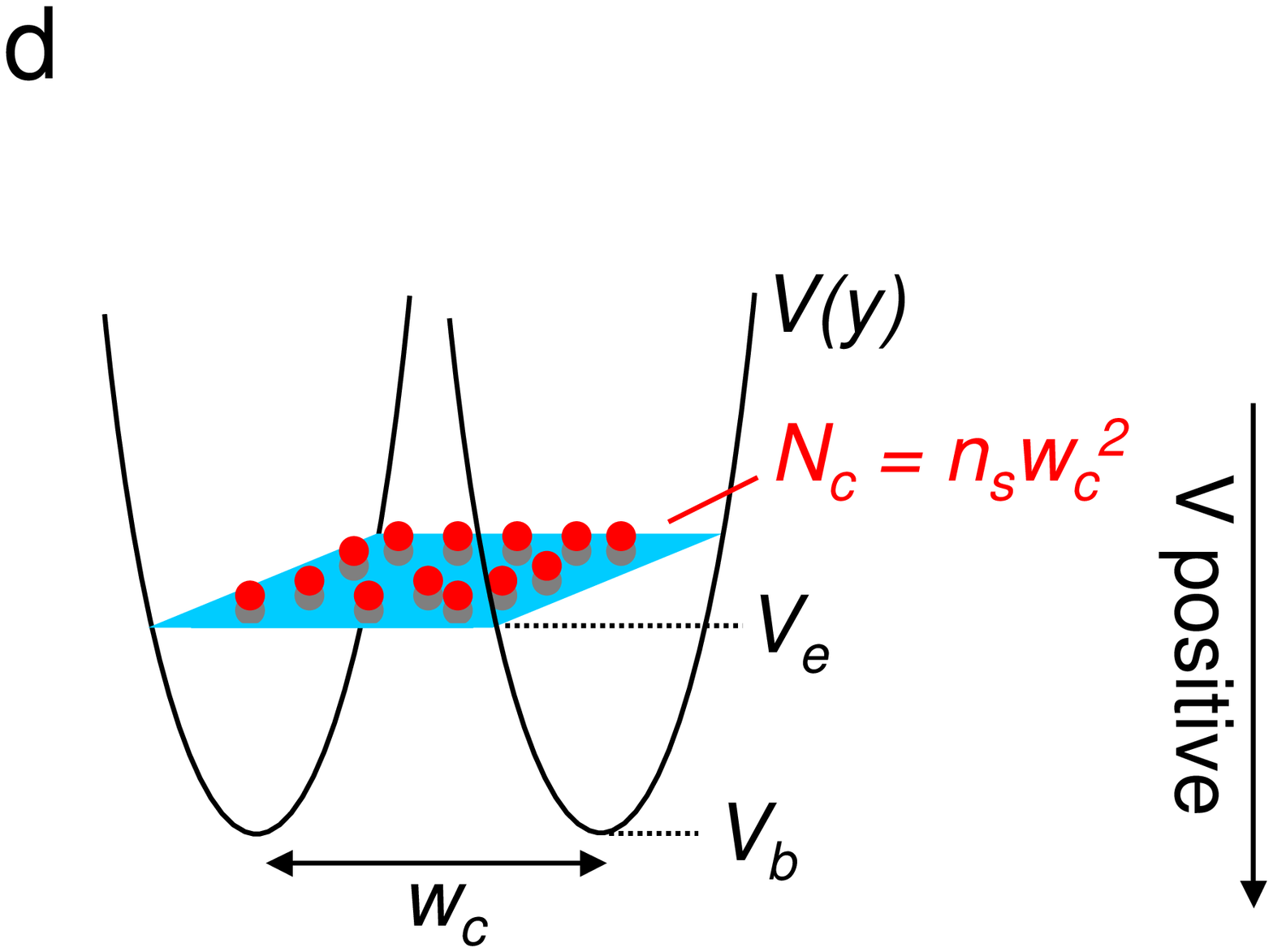}
\caption{(Color online) Two electrostatic models of the system. (a) Schematic diagram of the electrostatic potential across the device through the central channel, $V(x,0)$, for the case where $V_e=V_b$, assuming that the electron system forms a charge continuum. (b) As $V_r$ becomes more negative, $V_e$ becomes negative more quickly than $V_b$ and $V_{gt}$ must be set more negative to `pinch-off' the current. (c) Eventually $V_e=V_{gu}$ and when $V_r$ becomes more negative electrons are lost to the guard electrode. Because $V_e$ now remains constant, $V_{gt}$ must be set more positive to maintain the threshold condition. (d) Schematic representation of the constriction under a granular charge model in which the constriction is modeled as a small square of the helium surface of area $A=w_c^2$. The number of electrons in the constriction $N_c=n_sw_c^2$. Under this model, the current passing through the constriction should be suppressed when $N_c<1$.}
\label{fig:4}
\end{figure}

Before discussing experimental results, we first develop an electrostatic model of the device, following the approach presented previously\cite{Reesprl}. Note that in this model, a more positive voltage
corresponds to a lower potential energy for an electron. We estimate that the change in the depth of the helium at the center of the 20 $\mu$m-wide reservoir microchannels should be less than 0.1 $\mu$m. Therefore we do not take the curvature of the helium surface into account here. We begin by assuming that the electron system may be considered as a charge continuum. The electrostatic potential energy of the electron system in the reservoirs, $V_e$, depends on the reservoir electrode voltage $V_r$ and the electron density. Therefore we may write $V_e=-en_sd/\varepsilon\varepsilon_0+V_{r}$. The results of the finite element
modeling analysis (Fig. \ref{fig:3}) indicate that a saddle-point potential is formed at the constriction between the two reservoirs when $V_{gt}$ is more negative than $V_r$. At appropriate values of $V_{gt}$ a potential barrier between the two reservoirs may be formed. The saddle-point potential may be written as $V(x,y) = V_b+\frac{1}{2}ax^2-\frac{1}{2}by^2$ where $V_b$ is the potential at the center of the saddle-point and $a$ and $b$ are constants. We consider that for $-eV_e>-eV_b$ electrons may pass over the potential barrier, allowing transport through the constriction, whereas for $-eV_e<-eV_b$ electron transport is blocked. The condition $-eV_e=-eV_b$ therefore defines the threshold of current flow through the constriction. This condition is depicted schematically in Fig. \ref{fig:4}(a).

The influence of the reservoir, split-gate and guard electrodes on
the potential barrier may be estimated by considering the region
of space at the center of the constriction to have some
capacitance to the three electrodes, $C_{r}$, $C_{gt}$ and
$C_{gu}$, as well as a stray capacitance to the surroundings
$C_{s}$. The total
capacitance of the region is then described by $C_{\Sigma}= C_{r}+
C_{gt}+C_{gu}+C_{s}$ and we can define constants to represent the
relative strength of coupling from each electrode to the barrier
region as $\alpha=C_{r}/C_{\Sigma}$, $\beta=C_{gt}/C_{\Sigma}$,
$\gamma=C_{gu}/C_{\Sigma}$ and $\sigma=C_s/C_{\Sigma}$. Experimentally it was found that, on
increasing $V_r,V_{gt}$ and $V_{gu}$ by +100 mV simultaneously the threshold of current flow also
increased by exactly +100 mV (data not shown), indicating that
$C_{s}$ was in fact negligibly small ($\sigma=0$) and that
$\alpha+\beta+\gamma=1$. The potential at the center of the constriction may then be written as $V_b =
\alpha V_r + \beta V_{gt} + \gamma V_{gu}$.

We now consider the case in which the
reservoir voltage is set progressively more negative with a fixed
bias applied to the split gate and guard electrode (here we assume $V_{in}$
to be small). This is depicted schematically in Fig. \ref{fig:4}(a-c); from top to bottom $V_r$ goes from positive to negative
bias. In Fig. \ref{fig:4}(a) we see that, at a certain positive reservoir bias, the height
of the potential barrier just meets the potential energy of the
electron system and the current is `pinched-off'. Making $V_{r}$
more negative (Fig. \ref{fig:4}(b)) causes both $V_e$ and $V_{b}$ to become more
negative. As $V_e$ remains below the guard potential electrons do
not escape from the channels onto the thin helium film above the
guard; there is no change in the electron density and $V_e$ simply
changes by the change in $V_{r}$. However, the change in the
barrier height is smaller due its additional dependence on the split-gate and guard electrodes ($\alpha <1$). Therefore a negative change in
$V_{r}$ causes $V_{b}$ to become more positive relative to $V_e$ and current
flows across the barrier. By setting $V_{r}$ more negative still
(Fig. \ref{fig:4}(c)) electrons are eventually lost to the guard leading to a
reduction in $n_s$ until $V_e$ is equal to the guard potential.
Now as $V_{r}$ becomes more negative, $V_b$ becomes more negative with respect to $V_e$ and the current may eventually be suppressed once more.

We denote the value of $V_{gt}$ for which $-eV_e=-eV_b$ as $V^0_{gt}$. Following the model described above, expressions may now be derived to describe the dependence of $V^0_{gt}$ on $V_r$, for constant $V_{gu}$. From the threshold condition $-eV_e=-eV_b$, for the case where $-eV_e < -eV_{gu}$ we have
\begin{equation}
\frac{-en_sd}{\varepsilon\varepsilon_0} + V_r = \alpha V_r + \beta
V^0_{gt} + \gamma V_{gu}~~, \label{eq:1}
\end{equation}
and for the case where $-eV_e = -eV_{gu}$,
\begin{equation}
V_{gu} = \alpha V_r + \beta V^0_{gt} + \gamma V_{gu}~~. \label{eq:2}
\end{equation}
Rearranging (1) and (2) gives respectively,
\begin{equation}
V^0_{gt} = \frac{1-\alpha}{\beta} V_r -
\frac{\frac{en_sd}{\varepsilon\varepsilon_0} + \gamma
 V_{gu}}{\beta}~~, \label{eq:3}
\end{equation}
and
\begin{equation}
V^0_{gt} = \frac{-\alpha}{\beta} V_r + \frac{1 - \gamma
 }{\beta}V_{gu}~~. \label{eq:4}
\end{equation}

Because the separation between the split-gate electrodes (2.8 $\mu$m) is comparable to the inter-electron spacing ($\sim0.3$ $\mu$m for $n_s=1\times10^9$ cm$^{-2}$), we expect only a small number of electrons to be in the constriction region when the system is close to the threshold condition. To estimate this number, we model the center of the constriction as a small square of area $A=w_c^2$, where $w_c$ is the effective width of the constriction. As the FEM calculation indicates that the constant $a$ is small compared to $b$, we will assume that the potential in the $x$ direction is flat, whilst the electrons are confined in the $y$ direction by a parabolic potential $V(y)=V_b-\frac{1}{2}by^2$, as pictured schematically in Fig. \ref{fig:4}(d). The maximum lateral displacement for electrons in the constriction, $y_{max}$, satisfies the expression $V(y_{max})=V_b-\frac{1}{2}by_{max}^2=V_e$. Substituting $y_{max}=w_c/2$ gives
\begin{equation}
w_c =\sqrt{\frac{8}{b}(V_b-V_e)}~~. \label{eq:5}
\end{equation}
Note that for the threshold condition $-eV_e=-eV_b$, $w_c=0$. Assuming a parallel-plate capacitor approximation, the charge density in the area $A$ may be written as
\begin{equation}
n_c =\frac{\varepsilon\varepsilon_0}{ed}(V(y)-V_e)~~, \label{eq:6}
\end{equation}
and the total number of electrons in $A$ is therefore
\begin{equation}
N_c =\frac{\varepsilon\varepsilon_0}{ed}\int_{\frac{-w_c}{2}}^{\frac{w_c}{2}}\int_{\frac{-w_c}{2}}^{\frac{w_c}{2}}(V(y)-V_e)~dxdy~~. \label{eq:7}
\end{equation}
After integration over the limits as given by Eq. (\ref{eq:5}), we obtain the result
\begin{equation}
N_c=\frac{2}{3}\frac{8\varepsilon\varepsilon_0}{bed}(V_b-V_e)^2~~.
\label{eq:8}
\end{equation}

Although for the electron liquid electrons are not localized as in the Wigner crystal, we assume that electrons are distributed evenly over $A$, effectively forming a series of rows across the constriction. The number of electrons lying in the $y$ direction across the constriction may then be estimated as $N_y=\surd N_c$. For constant $V_r$ and $V_{gu}$, we also have the relation $V_b-V_e=\beta(V_{gt}-V^0_{gt})$. This gives an expression relating $N_y$ to $V_{gt}$ as
\begin{equation}
N_y=\sqrt{\frac{2}{3}\frac{8\epsilon\epsilon_0}{bed}}\beta(V_{gt}-V^0_{gt})~~.
\label{eq:9}
\end{equation}
The number of electrons across the constriction should therefore increase linearly with increasing $V_{gt}$ above the current threshold.

We suggest that the granularity of charge may cause some deviation from the behavior of the system as expected under the charge continuum model. From Eq. (\ref{eq:9}), the value of $V_{gt}$ for which one row of electrons may be formed in the constriction ($N_y=1$) is
\begin{equation}
V^1_{gt}=V^{0}_{gt}+\frac{1}{\beta}\sqrt{\frac{3}{2}\frac{bed}{8\epsilon\epsilon_0}}~~.
\label{eq:10}
\end{equation}
For $V^0_{gt}\leq V_{gt}< V^1_{gt}$ the number of electron rows across the constriction is less than 1. Therefore, as no electrons are present in the constriction, current flow should be suppressed when $V_{gt}=V^1_{gt}$ rather than at the threshold condition assumed under the charge continuum model $V_{gt}=V^0_{gt}$. We may now correct Eqs. (\ref{eq:3}) and (\ref{eq:4}) to give $V^1_{gt}$ for changing $V_r$ at constant $V_{gu}$. For the case where $-eV_e < -eV_{gu}$ we have
\begin{equation}
V^{1}_{gt} = \frac{1-\alpha}{\beta} V_r +\frac{1}{\beta}\sqrt{\frac{3}{2}\frac{b(V_r)ed}{8\epsilon\epsilon_0}}- \frac{\frac{en_sd}{\varepsilon\varepsilon_0} + \gamma V_{gu}}{\beta}~~, \label{eq:11}
\end{equation}
and for the case where $-eV_e = -eV_{gu}$,
\begin{equation}
V^{1}_{gt} = \frac{-\alpha}{\beta} V_r +\frac{1}{\beta}\sqrt{\frac{3}{2}\frac{b(V_r)ed}{8\epsilon\epsilon_0}}+ \frac{1 - \gamma}{\beta}V_{gu}~~. \label{eq:12}
\end{equation}
Note that the value of $b$, which describes the parabolic lateral confinement at the constriction, is dependent on the potential applied to the reservoir electrodes (assuming constant $V_{gu}$). The FEM calculation of the potential profile shows this dependency to be linear. The corrective terms therefore introduce a non-linear dependence of $V^{1}_{gt}$ on $V_r$. However, in the following discussion, we will assume that, for constant $V_r$, small changes in $V_{gt}$ cause a negligibly small change in $b$.

\begin{figure}
  \includegraphics[angle=0,width=0.45\textwidth]{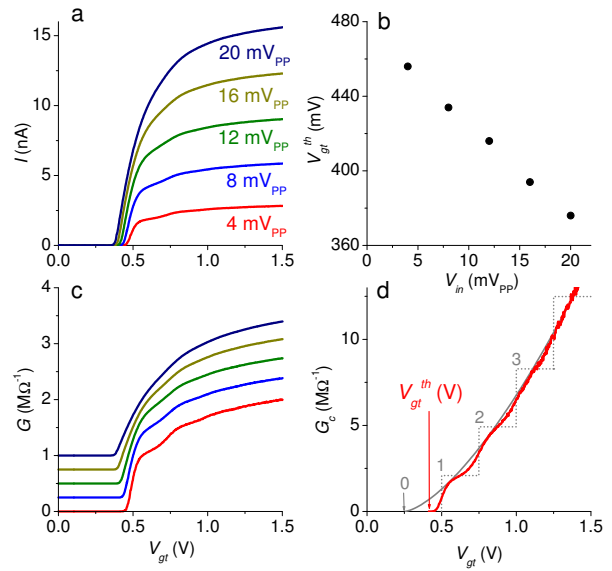}
\caption{(Color online) (a) Peak current $I$ measured as a function of $V_{gt}$ for different driving voltage amplitude $V_{in}$. (b) The current threshold $V^{th}_{gt}$ against $V_{in}$. (c) Conductance of the electron system $G$ as a function of $V_{gt}$ for different $V_{in}$. Above 4 mV$_{\mathrm{pp}}$ each data set is shifted vertically for clarity. (d) Constriction conductance $G_c$ for $V_{in}=4$ mV$_{\mathrm{pp}}$. $G_c$ increases in a series of smoothed steps. The grey dotted line is a guide to the eye indicating values of $V_{gt}$ at which, under the granular charge model, the number of electrons across the constriction increases by 1. The threshold $V^{th}_{gt}$ lies between $N_y=0$ and $N_y=1$. The solid grey line is a guide to the eye of functional form $G_c\sim V_{gt}^{\frac{3}{2}}$ starting at $V_{gt}=0.25$ V.}
\label{fig:5}
\end{figure}

We have performed measurements in order to determine whether the charge continuum model or the granular charge model best describes the dynamics of electrons in the device. We denote the experimentally determined value of the split-gate voltage for which the current is suppressed as $V^{th}_{gt}$. Figure \ref{fig:5}(a) shows the dependence of the current
\emph{I} flowing through the central channel as $V_{gt}$ is varied
for different values of the driving voltage $V_{in}$. The current
decreases as $V_{gt}$ is swept negative and finally is completely
suppressed at the threshold voltage $V^{th}_{gt}$, which for all
cases is more negative than the reservoir voltage $V_{r} = + 1$ V.
The linear dependence of $V^{th}_{gt}$ on $V_{in}$ is shown in
Fig. \ref{fig:5}(b). The threshold for current flow occurs at more negative
split-gate bias as the driving voltage is increased.

Measurements of SSE current flow over
a potential barrier formed by a split-gate electrode in a similar device
have already been reported\cite{ourJLTP}. There, for values of $V_{in}$ up to
100 mV$_{\mathrm{pp}}$, it was shown that current flow across a potential
barrier could be induced by increasing $V_{in}$ until the
potential energy of the electron system could overcome the barrier,
allowing electrons to be transferred between the reservoirs. Here
we observe the same behavior. The dependence of $V^{th}_{gt}$ on $V_{in}$ is discussed in more detail later in this section.

In Fig. \ref{fig:5}(c) we show the corresponding conductance \emph{G} for
each current measurement. Above $V^{th}_{gt}$, \emph{G} increases
sharply for all values of $V_{in}$. Considering the saddle-point
potential shown in the modeling results
in Fig. \ref{fig:3}, both the depth of the potential at the center of the constriction (which is proportional to the electron density in the constriction under the charge continuum model), and the
effective width $w_c$ of the constriction, should increase as
$V_{gt}$ is swept positive. Both effects should lead to an increase of conductance at the constriction. For
high values of $V_{gt}$, where the split-gate electrode is more
positive than the reservoir electrode and no potential barrier is
expected to exist, \emph{G} rises further, presumably due to the
continuing increase of the electron density in the constriction
region.

For still higher values of $V_{gt}$, the conductance eventually saturates. For $V_{in}=4$ mV$_{\mathrm{pp}}$, $G=2.30$ M$\mathrm{\Omega}^{-1}$ at $V_{gt}=3.0$ V. As the total conductance of the electron system no longer depends on the split-gate voltage in this region, we assume that the resistance of the constriction region is small compared to that of the electron system in the reservoirs. Therefore, the saturated conductance can be attributed to that of the reservoir region, and by subtracting the value of the resistance at this split-gate voltage, $R = 1/G = 0.434$ M$\mathrm{\Omega}$, from the values of the resistance over the entire split-gate sweep, an approximation of the constriction resistance $R_c$ can be calculated. The corresponding conductance of the constriction, $G_c=1/R_c$, is shown in Fig. \ref{fig:5}(d), for $V_{in}=4$ mV$_{\mathrm{pp}}$. We see that above the conductance threshold, $G_c$ increases in a series of steps. The steps are not a series of sharp rises and flat plateaus; rather, they appear smoothed-out. We add to the plot a guide to the eye (grey dotted line) indicating the manner in which a sharp step pattern fits the data. We align each sharp step with the maximum in the gradient at each increase in $G_c$. The spacing between the steps is $\Delta V_{gt}=250$ mV.

We suggest that the increasing number of electron rows across the constriction could lead to steplike increases in the constriction conductance. Such behavior is observed in other classical many-body systems with long-range interactions, such as pedestrians moving through bottlenecks\cite{Hoogendoorn2005}. Here, such an effect is essentially the result of Coulomb blockade at a single constriction; for the case in which $N_y=1$ and one electron row occupies the constriction, Coulomb repulsion prevents other electrons from passing through and electrons may only pass through in a single row. As the constriction is opened to the point where $N_y=2$, the additional conduction channel should cause an increase
in the constriction conductance as electrons may now pass freely through the constriction, side-by-side. The steplike increase in conductance resembles the quantized conductance steps observed in quantum point contact devices\cite{OriginalQPC}. From the FEM calculation results shown in Fig. \ref{fig:3}(b), we estimate that the spacing of the energy subbands for lateral motion at the constriction to be $\sim0.1$ meV, which, as shown below, is much smaller than the change in $V_b$ associated with each conductance step, and is also smaller than $V_{in}$. We therefore conclude that the origin of the steps observed here, due to the Coulomb interaction between electrons at the constriction, is quite different to the case of the quantum point contact, of which our device acts as a classical analogue.

The smoothing of the steps in $G_c$ indicates that neither the charge continuum model, nor the granular charge model, describes the system precisely. We assume that the sharp steps fitted to the data in Fig. \ref{fig:5}(d) correspond to the values of $V_{gt}$ where, under the granular charge model, the number of electrons able to pass simultaneously through the constriction increases by 1. By extrapolation of these fitted steps, we can obtain an estimate of $V^0_{gt}=0.25$ V. As described above, the constriction conductance should increase with increasing $w_c$ and as $V_b$ becomes more positive. The FEM calculation shows that $w_c\sim \surd V_{gt}$ whereas $V_b\sim V_{gt}$. We therefore naively expect the conductance of the constriction to vary as $G_c\sim V_{gt}^{\frac{3}{2}}$, under the charge continuum model. In Fig. \ref{fig:5}(d) we plot the function $G_c=1.04\times10^7 (V_{gt}-V^0_{gt})^{\frac{3}{2}}$ (solid grey line). We see that this function describes $G_c$ reasonably well, other than at points along the curve where the conductance is suppressed, which we attribute to deviations from the continuum model due to the granular nature of charge. The agreement becomes closer for higher values of $V_{gt}$ as the step features are lost, presumably as the number of electron rows increases and the electron system at the constriction better approximates a 2D charge continuum. We therefore conclude that whilst the granularity of charge causes observable deviations from the charge continuum model, and causes the current to be suppressed above the expected threshold $V^0_{gt}$, the simplified approach depicted in Fig. \ref{fig:4}(d) does not describe the system accurately. Correspondingly, the experimentally observed threshold of current flow, $V^{th}_{gt}=0.456$ V, lies between the values which we have estimated correspond to $V^0_{gt}$ and $V^1_{gt}$.

We suggest that our experimental observation of the step-like increase in the conductance of the constriction may be verified by molecular dynamics simulations of classical charge systems at potential bottlenecks\cite{PeetersConstriction}. Indeed, recent simulations of a system very similar to ours reproduce the smooth steplike increase in $G_c$, and show that each step is related to an increase in the number of electron rows across the constriction\cite{Araki2011}. These simulations also indicate that temporal fluctuations in the potential of electrons at the constriction, which are of thermal origin but essentially due to electron-electron interactions, and are not considered in our mean-field approach, cause the smoothing of the conductance steps. These fluctuations also cause $V^{th}_{gt}$ to lie at vales of $V_{gt}$ more negative than $V^1_{gt}$ as, even for the case $N_y<1$, there still exists some probability that the potential of electrons may be raised in order to overcome the barrier at the constriction, thus allowing transport. For our experiments, the driving voltage $V_{in}$ causes an additional modulation of the electron density at the constriction over each AC cycle. In Fig. \ref{fig:5}(c) the smoothing of the step-like features increases with increasing $V_{in}$, as discussed in more detail later in this section. However, we note here that extrapolation of the data shown in Fig. \ref{fig:5}(b) to the limit $V_{in}=0$ mV$_{\mathrm{pp}}$ yields a value of $V^{th}_{gt}\approx0.47$ V, which is a small change from the value recorded for $V_{in}=4$ mV$_{\mathrm{pp}}$ and is still more negative than the value of $V^1_{gt}=0.5$ V. Further comparison of our experimental results with numerical simulations is required to fully understand the dynamics of the electron system close to the conductance threshold.

\begin{figure}
  \includegraphics[angle=0,width=0.23\textwidth]{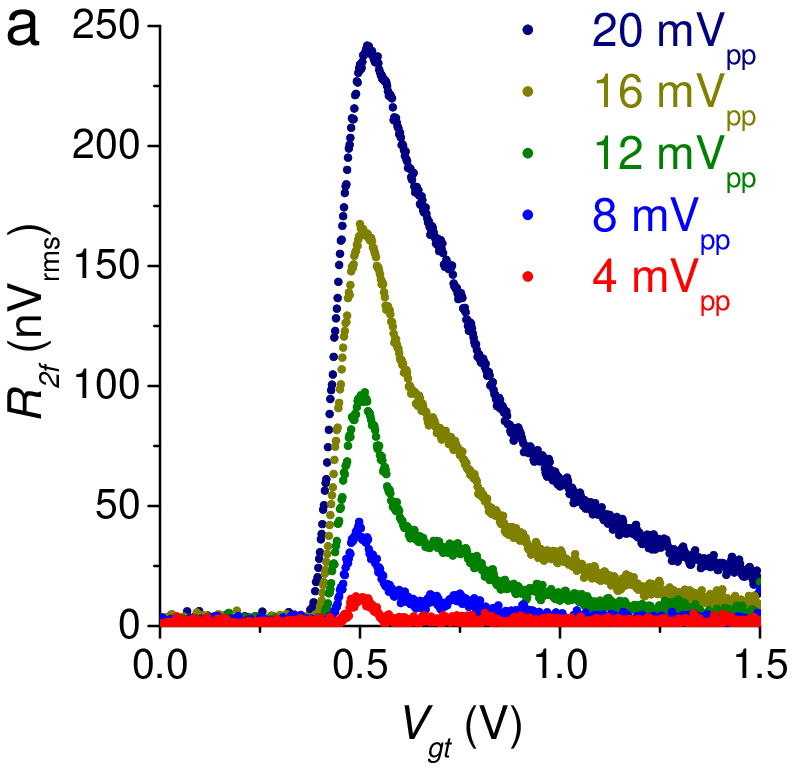}
    \includegraphics[angle=0,width=0.23\textwidth]{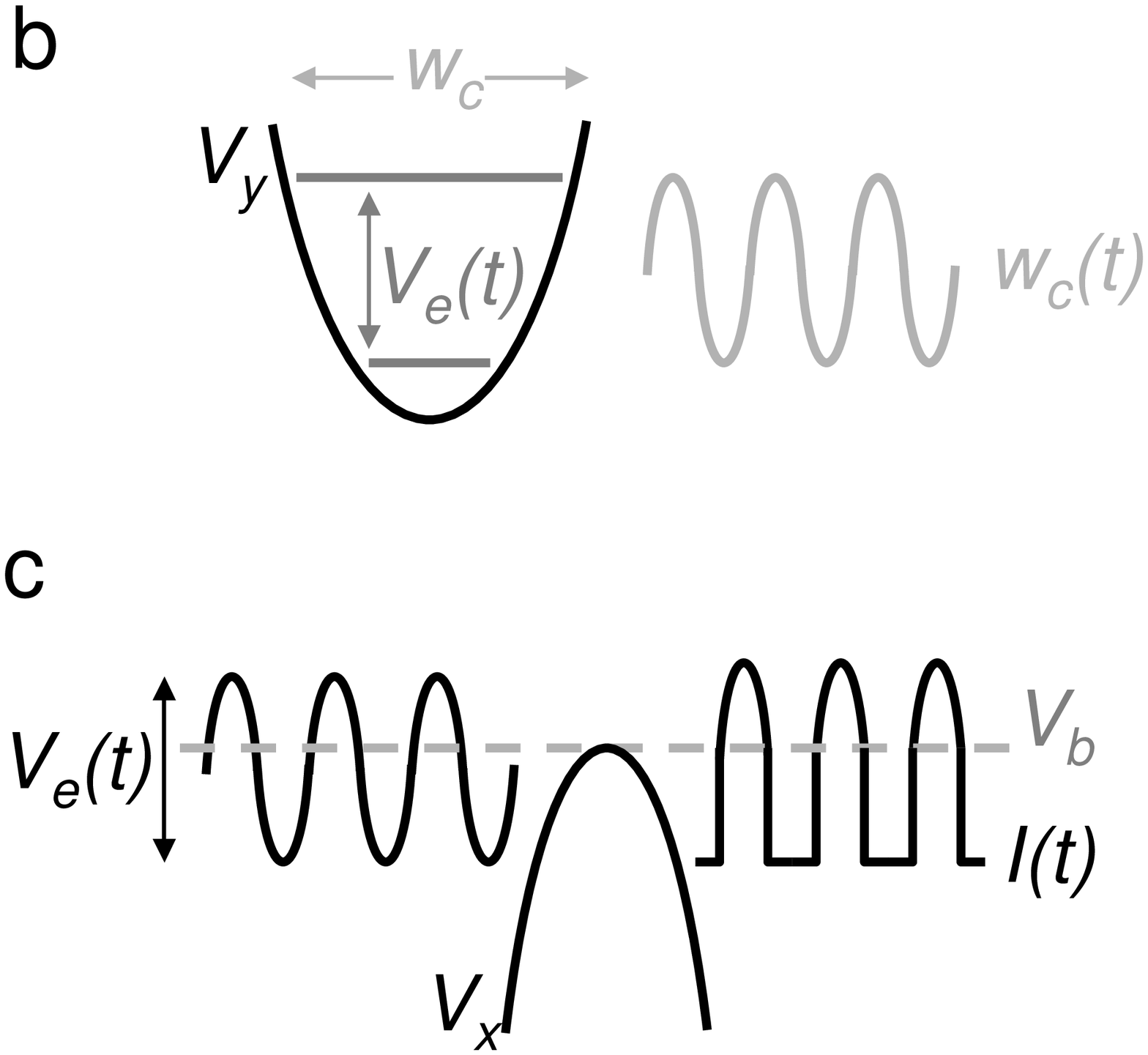}
\caption{(Color online) (a) Magnitude of the second harmonic component of the voltage signal, $R_{2\!f}$, as a function of $V_{gt}$ for different driving voltages. (b) Processes giving rise to distortion in the AC current through a saddle-point potential. As the potential
across the constriction $V_y$ is parabolic, the effective width of the constriction $w_c$ oscillates over each AC cycle of $V_{in}$, causing the conductance of the constriction region to vary over each cycle. (c) Close to the threshold of
current flow, electrons may only pass intermittently across the potential barrier formed at the constriction.}
\label{fig:6}
\end{figure}

We consider that the modulation of the potential of the electron
system due to the applied driving voltage may cause additional
non-linear behavior to appear in the electron transport properties
close to the current threshold. To investigate such effects, the
magnitude of the second harmonic component of the AC voltage induced on the left reservoir electrode, $R_{2f}$, was measured. In the case that the SSE AC
current flow is perfectly sinusoidal in response to the driving
voltage, the second harmonic component is zero. However, if
the conductance should vary over each AC cycle, the
current signal should become distorted causing higher frequency
components to appear. In Fig. \ref{fig:6}(a) we see that $R_{2\!f}$ indeed
rises as $V_{gt}$ is swept negative and reaches a maximum close to
$V^{th}_{gt}$. The magnitude of $R_{2\!f}$ increases with the
driving voltage $V_{in}$.

In Fig. \ref{fig:6}(b) and (c) we depict schematically two processes which may give
rise to the distortion of the electron current through a
saddle-point potential as the electron energy $V_e$ is modulated.
In the first case (Fig. \ref{fig:6}(b)), when $V_e$ is close to the bottom
of the parabolic potential $V_y$, the effective width of the
conductive channel $w_c$ changes over the AC cycle which should
lead to a time dependence of the conductance \emph{G}. In addition, the effective depth of the potential for electrons at the constriction also varies, which should lead to a variation of the electron density, and so conductance, over each cycle. In the
second case (Fig. \ref{fig:6}(c)), as the potential maximum in $V_x$ is
raised and becomes higher than $V_e-V_{in}/2$, electrons are
expected to flow only intermittently across the barrier.

In both of these
cases, the degree of distortion in the current should reach a
maximum close to the threshold of current flow. We therefore find the
observed increase in $R_{2\!f}$ close to the current threshold to
be consistent with transport through a saddle-point
potential. We also note that $R_{2\!f}$ appears to rise in a series of weak peaks or steps which appear to mirror the step-like decrease in conductance observed in Fig. \ref{fig:5}(c). We again find this to be consistent with our model. For values of $V_{gt}$ where $dI/dV_{gt}$ is large, modulating $V_e$ gives rise to a large change in conductance over each AC cycle, and so the value of $R_{2\!f}$ should be large. Each step in conductance should therefore be accompanied by a peak in $R_{2\!f}$. Indeed, the step-like feature in the conductance at $V_{gt}=0.75$ V (Fig. \ref{fig:5}(c)) is accompanied by a weak second peak in $R_{2\!f}$ at the same split-gate voltage. However, the non-linear response
of the SSE system to a driving field has been predicted by Saitoh\cite{saitoh}
and observed experimentally\cite{bridges}. We therefore note that the distortion in the current signal may also be due in some part to
the intrinsic non-linear transport properties of the SSE system rather than solely due to
the geometry or potential profile of the device.

\begin{figure}
  \includegraphics[angle=0,width=0.3\textwidth]{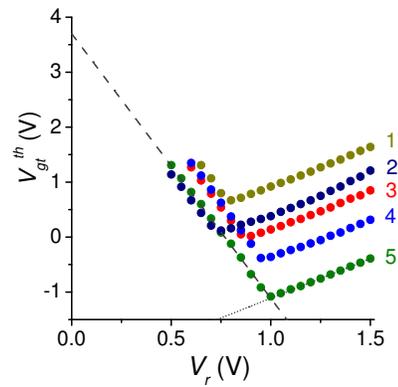}
\caption{(Color online) Measured values of $V^{th}_{gt}$ for different values of $V_r$, which was varied from positive to negative values, and $V_{in}=8$ mV$_{\mathrm{pp}}$. The measurement was performed five times, as indicated by the different colors. The helium surface was charged at the start of each measurement. The dotted and dashed lines correspond to Eqs. (\ref{eq:3}) and (\ref{eq:4}) respectively. As explained in the text, for the data sets 1-5, $n_s= 0.70, 0.50, 0.99, 1.28, 1.48\times10^{9}$ cm$^{-2}$ respectively.}
\label{fig:7}
\end{figure}

The dependence of $V^{th}_{gt}$ on $V_r$ was investigated by measuring the
current threshold for decreasing values of $V_r$, from 1.5 V to 0 V in 50 mV steps, for $V_{in}=8$ mV$_{\mathrm{pp}}$. The results of 5 such measurements, taken on different days,
are shown in Fig. \ref{fig:7}. In all 5 cases the split-gate voltage
required to suppress the current initially becomes more negative
as the reservoir potential is made more negative. Then, at a
certain reservoir potential, the trend is reversed; the threshold
moves to more positive values as the reservoir potential becomes
more negative. In both cases the relationship between
$V^{th}_{gt}$ and $V_{r}$ is approximately linear. These results
are found to be in agreement with the behavior predicted by our electrostatic analysis. Starting at
highly positive reservoir electrode bias, $V_{gt}$ must initially be made
more negative to suppress the current flow each time $V_r$ is set more negative, to compensate for the
reduction in the barrier height relative to $V_e$. Then, as electrons are lost to the
guard, $n_s$ decreases and $V_e$ remains constant, the split-gate
voltage must be made more positive to allow current flow with each step in $V_r$, due to
the relative increase of the barrier height. The intersection between the two linear regions in the
data marks the point at which electrons begin to escape to the
guard. This point is different for each data set indicating that
the initial surface density in each case was different. The
intersection was not observed to occur for values of $V_{r}$
greater than +1.0 V, the voltage at which the electron reservoirs
were charged.

By making linear fits to the data presented in Fig. \ref{fig:7}, we may derive values for $\alpha$, $\beta$ and $\gamma$ based on the charge continuum model, using Eqs. (\ref{eq:3}) and (\ref{eq:4}), and the relation $\alpha+\beta+\gamma=1$. The values of the
coupling constants were determined for each of the data sets. The
similarity in the gradient of each data set indicates the
stability of the coupling constants, the average values of which
are shown in Table \ref{tab:1}.

We may also estimate $\alpha$, $\beta$ and $\gamma$ taking into account the fact that the experimentally determined current threshold does not correspond exactly to $V^0_{gt}$. The results in Fig. \ref{fig:5} show that $V^{th}_{gt}$ appears to lie closer to $V^1_{gt}$, the threshold under the granular charge model, as given by Eqs. (\ref{eq:11}) and (\ref{eq:12}).  Whilst the constant $b$, which determines the deviation of $V^0_{gt}$ from $V^1_{gt}$, may be estimated from the FEM modeling of the device, here we adopt a more straight-forward approach to take into account the deviation of $V^{th}_{gt}$ from $V^0_{gt}$. From Fig. \ref{fig:5}, for $V_r=1.0$ V, $V^{th}_{gt}-V^0_{gt}=0.16$ V. As discussed in more detail below, the reservoir electrode potential is equal to the guard electrode potential for $V_{gu}\approx 0.5$ V, due to an offset in the potential of the guard electrode. Because no parabolic confinement exists when $V_r=V_{gu}$, for this condition we expect $b=0$ and therefore $V^{th}_{gt}-V^0_{gt}=0$ V. As the correction to $V^0_{gt}$ should display a square-root dependence on $V_r$, the value of $V^0_{gt}$ can be estimated for each value of $V_r$ using the expression
\begin{equation}
V^{0}_{gt} = V^{th}_{gt}- 0.16\sqrt{\frac{V_r-0.5}{1-0.5}}~~. \label{eq:13}
\end{equation}
The average values for $\alpha$, $\beta$ and $\gamma$ calculated using the values for $V^0_{gt}$ given by Eq. (\ref{eq:13}) are also listed in Table \ref{tab:1}.

The coupling constant $\beta$ may also be estimated from dependence of $V^{th}_{gt}$ on $V_{in}$ shown in Fig. \ref{fig:5}(b). We assume that, on increasing
$V_{in}$, the increase in the barrier height required to suppress
the current is equal to the corresponding increase in the maximum
energy of the electron system in the right reservoir, $V_{in}/2$.
Because, for fixed bias on the reservoir and guard electrodes, $\Delta V_b=\beta \Delta V_{gt}$ we may derive $\beta$ from the gradient in
Fig. \ref{fig:5}(b) as $d V^{th}_{gt}/d V_{in} = -1/2\beta$.
Similar measurements were made by sweeping the
reservoir and guard potentials (data not shown) in order to determine the
dependence of the thresholds $V^{th}_{r}$ and $V^{th}_{gu}$ on
$V_{in}$. Under the charge continuum model, the constants
$\alpha$ and $\gamma$ may be given by $d
V^{th}_{r}/d V_{in} = 1/2(1-\alpha)$ and $d
V^{th}_{gu}/d V_{in} = -1/2\gamma$. The results of these
measurements (denoted $dV^{th}/d V_{in}$) are shown in Table \ref{tab:1} as well as the corresponding
values calculated from the finite element model.
\begin{table*}[!h]
\caption{\label{tab:1}Coupling constants $\alpha, \beta$ and $\gamma$ as calculated by the finite element analysis and the values measured experimentally from the results shown in Fig. \ref{fig:5} and Fig. \ref{fig:7}.}
\begin{ruledtabular}
\begin{tabular}{lllll}
\hline\noalign{\smallskip}
Coupling constant & Calculated (FEM) & Measured (Eqs. (\ref{eq:3}), (\ref{eq:4})) & Measured (Eq. (\ref{eq:13})) & Measured ($dV^{th}/dV_{in}$) \\
\noalign{\smallskip}\hline\noalign{\smallskip}
$\alpha$ & 0.75 & 0.77 & 0.79 & 0.83 \\
$\beta$ & 0.10 & 0.16 & 0.15 & 0.10 \\
$\gamma$ & 0.15 & 0.07 & 0.06 & 0.07 \\
\noalign{\smallskip}\hline
\end{tabular}
\end{ruledtabular}
\end{table*}

From the results shown in Table \ref{tab:1}, we see that, compared to the values calculated using Eqs. (\ref{eq:3}), (\ref{eq:4}), the change in $\alpha$,
$\beta$ and $\gamma$ when including the correction given by Eq. (\ref{eq:13}) is small. These values are also close to those given by measuring the dependence of $V^{th}$ on $V_{in}$. All the experimentally determined values are in relatively good agreement with the FEM calculation. For all four results, $\alpha$ is largest and so the
reservoir electrode dominates in determining the height of the
potential barrier. This confirms that in our experiment, close to the conductance threshold, the electrons are indeed passing through
the region between the split-gate electrodes and above the
reservoir electrode.

The determination of the constant $\beta$ allows the expected change in $V_{gt}$ to add one electron row across the constriction to be calculated using Eq. (\ref{eq:10}). We use the FEM model to obtain an estimate of $b=2.5\times10^{11}$ Vm$^{-2}$ for $V_r=1.0$ V and $V_{gu}$=0.62 V. Using $\beta=0.16$ we find that $V^1_{gt}-V^{0}_{gt}=224$ mV. This value is in good agreement with the experimentally observed step separation, $\Delta V_{gt}=250$ mV for $V_r=1.0$ V (Fig. \ref{fig:5}(d)), indicating that the step-like increases in $G_c$ are indeed due to increases in the number of electrons across the constriction. The change in split-gate voltage $\Delta V_{gt}$ is related to the corresponding change in $V_b$ by the constant $\beta$. For $\beta=0.16$, $\Delta V_b\approx40$ mV, which is comparable to $V_{in}$. We therefore conclude that the increase of smoothing of the steplike features with increasing $V_{in}$ may be due to the modulation of the electron density at the constriction, which causes transport features arising from the discreet number of electrons across the constriction to be lost.

The data shown in Fig. \ref{fig:7} was taken with a voltage of $0$ V applied to the guard electrode. From
Eq. (\ref{eq:4}) it is then to be expected that $V^{th}_{gt} = 0$ V for
$V_r = 0$ V. However, the intercept on the
$V^{th}_{gt}$ axis for each data set is typically +3.6 V
indicating a true value of the guard potential of $V_{gu} \approx
+0.62$ V. This apparent offset is confirmed experimentally for
each of the data sets; below a value of $V_r \approx 0.5$ V the electron
signal was lost completely indicating that at this point $V_r
\approx V_{gu}$. The offset is seen to vary over a small range;
for data sets with a higher intercept on the $V^{th}_{gt}$ axis
the electrons were lost at more positive values of $V_r$. Despite thermally
cycling the device to room temperature and making efforts to
ensure correct grounding of the experimental wiring the offset
remained. However, the stability of the offset over many hours or days allowed
consistent measurements to be made. Voltage
offsets in mesoscopic devices may be caused by contact potential
differences, thermoelectric effects or surface charging effects\cite{mukharsky}. The cause of the offset observed in this experiment is not yet clear. Taking the offset into account, the estimated initial surface electron densities for the data sets 1-5 shown in Fig. \ref{fig:7} are 0.70, 0.50, 0.99, 1.28, $1.48\times10^{9}$ cm$^{-2}$ respectively.

\begin{figure}
  \includegraphics[angle=0,width=0.3\textwidth]{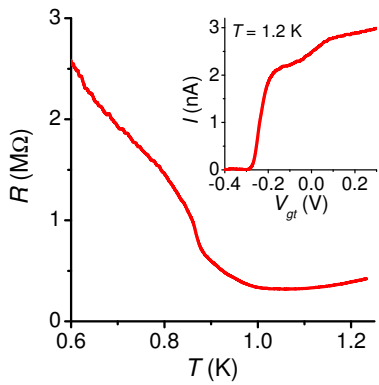}
\caption{(Color online) Resistance against temperature for the case in which the constriction is widely opened ($V_{gt}=+1$ V) for $V_{in}=5$ mV$_{\mathrm{pp}}$ . Inset: $I$ as a function of $V_{gt}$ for the same electron density, measured at 1.23 K.}
\label{fig:8}
\end{figure}

The melting temperature of the 2D Wigner solid depends on the
electron density as $T_m = 0.225\times 10^6 \surd{n_s}$\cite{Phil'sChannels}. For
the case of a potential offset on the guard potential of +0.62 V,
the saturated electron density achieved when charging the
reservoirs with $V_r = +1$ V is $n_s = 1.48\times10^{9}$ cm$^{-2}$
which gives a value of $T_m = 0.866$ K. In Fig. \ref{fig:8} we show the
temperature dependence of the resistance $R$ of the electron
system for $V_{r} = V_{gt} = +1.0$ V. At 1.23 K the current
threshold was measured (inset). In reference to Fig. \ref{fig:7}, the
highly negative value of the threshold $V^{th}_{gt} = -0.29$ V
indicates that the density was close to saturation. The increase
in $R$ below 1 K is attributed to the formation of the 2D Wigner
solid. The localization of electrons in the Wigner lattice leads to the formation of a small depression, or dimple, in the helium surface beneath each electron, which increases the electron effective mass, and so resistivity of the system\cite{ShikinMonarkhaDimples}. Recent experiments have investigated the transport of the Wigner solid in a microchannel geometry where the decoupling of the electron lattice from the dimple lattice at high driving fields leads to a highly non-linear response\cite{IkegamiWigner}. In our measurement, a sharp
increase in the resistance is observed at $T \approx 0.875$ K which may
correspond to the melting temperature of the electron system, in
agreement with the predicted value of $T_m$ for the case in which the
guard potential is offset.

\section{Conclusions}
\label{Conclusions} We have investigated the AC transport
of strongly-correlated electrons on the surface of liquid
helium at a constriction formed by a split-gate electrode.
The electron current may be suppressed by
sweeping the voltage of the split gate negative. The threshold for
current flow was dependent on the DC voltages of all the device
electrodes and the AC driving voltage applied to the electron system, as well as the electron density. Step-like increases in the conductance of the electron system as the split-gate voltage was swept positive were found to be due to increases in the number of electrons able to pass simultaneously through the constriction. The device therefore acts as a classical analogue of the quantum point contact. Our results are in
good agreement with a simple model of the device developed with
the aid of finite element analysis software in which a
saddle-point potential profile is formed at the constriction. Comparison with this model reveals that a potential offset on the guard electrode of the device plays a crucial role in determining the potential profile of the sample. Such detailed characterization of
microfabricated samples for electrons on the surface of liquid
helium is an important step towards the realization of more
advanced mesoscopic devices such as single electron devices and
quasi-one dimensional wires.
\begin{acknowledgements}
We thank M. Dykman, F. Nori, K. Ono, H. Totsuji, M. Araki and H. Hayakawa for useful discussions. This work was partially supported by Kakenhi. DGR was supported by the RIKEN FPR program.
\end{acknowledgements}

\bibliographystyle{spphys}       

\begin{thebibliography}{10}
\providecommand{\url}[1]{{#1}}
\providecommand{\urlprefix}{URL }
\expandafter\ifx\csname urlstyle\endcsname\relax
  \providecommand{\doi}[1]{DOI \discretionary{}{}{}#1}\else
  \providecommand{\doi}{DOI \discretionary{}{}{}\begingroup
  \urlstyle{rm}\Url}\fi

\bibitem{Andrei}
E.~Andrei (ed.), \emph{Two-dimensional electron systems on helium and other
  cryogenic substrates.} (Kluwer Academic, Dordrecht, 1997)

\bibitem{MonarkhaKono}
Y.P. Monarkha, K.~Kono, \emph{Two-Dimensional Coulomb Liquids and Solids}
  (Springer-Verlag, Berlin, 2004)

\bibitem{GrimesImageStates}
C.C. Grimes, et~al., Phys. Rev. B \textbf{13}(1), 140 (1976)

\bibitem{PlatzmanandDykmanscience}
P.M. Platzman, M.I. Dykman, Science \textbf{284}, 1967 (1999)

\bibitem{Gorkov}
L.~Gor'kov, D.~Chernikova, JETP Lett. \textbf{18}(2), 68 (1973)

\bibitem{Leiderer1979189}
P.~Leiderer, M.~Wanner, Phys. Lett. A \textbf{73}(3), 189  (1979)

\bibitem{LeidererDimples}
M.~Wanner, P.~Leiderer, Phys. Rev. Lett. \textbf{42}(5), 315 (1979)

\bibitem{ShirahamaMobility}
K.~Shirahama, S.~Ito, H.~Suto, K.~Kono, J. Low Temp. Phys. \textbf{101}, 439
  (1995)

\bibitem{GrimesAdamsWignerCrystal}
C.C. Grimes, G.~Adams, Phys. Rev. Lett. \textbf{42}(12), 795 (1979)

\bibitem{KovdryaReview}
Y.Z. Kovdrya, Low Temp. Phys. \textbf{29}(77) (2003)

\bibitem{KovdryaNikolaenko}
Y.Z. Kovdrya, V.A. Nikolaenko, Sov. J. Low Temp. Phys. \textbf{18}(894) (1992)

\bibitem{KovdryaNikolaenko2}
Y.Z. Kovdrya, et~al., J. Low Temp. Phys. \textbf{110}(1), 191 (1998)

\bibitem{Sokolov1Dmobility}
S.S. Sokolov, G.Q. Hai, N.~Studart, Phys. Rev. B \textbf{51}(9), 5977 (1995)

\bibitem{MartyCurvature}
D.~Marty, Journal of Physics C: Solid State Physics \textbf{19}(30), 6097
  (1986)

\bibitem{vanHaren}
R.~van Haren, et~al., Physica B: Condensed Matter \textbf{249-251}, 656  (1998)

\bibitem{Phil'sChannels}
P.~Glasson, et~al., Phys. Rev. Lett. \textbf{87}(17), 176802 (2001)

\bibitem{IkegamiWigner}
H.~Ikegami, H.~Akimoto, K.~Kono, Phys. Rev. Lett. \textbf{102}(4), 046807
  (2009)

\bibitem{sabouret}
G.~Sabouret, et~al., Appl. Phys. Lett. \textbf{92}(8), 082104 (2008)

\bibitem{CountingElectronsOnLHe}
G.~Papageorgiou, et~al., Appl. Phys. Lett. \textbf{86}(15) (2005)

\bibitem{heliumFET}
J.~Klier, I.~Doicescu, P.~Leiderer, J. Low Temp. Phys. \textbf{121}(5-6), 603
  (2000)

\bibitem{Chaplik}
A.V. {Chaplik}, Pis'ma Zh. Eksp. Teor. Fiz. \textbf{31}, 275 (1980)

\bibitem{Peeters1DCrystal}
G.~Piacente, I.V. Schweigert, J.J. Betouras, F.M. Peeters, Phys. Rev. B
  \textbf{69}(4) (2004)

\bibitem{bedanov}
V.M. Bedanov, F.M. Peeters, Phys. Rev. B \textbf{49}(4), 2667 (1994)

\bibitem{PeetersConstriction}
G.~Piacente, F.M. Peeters, Phys. Rev. B \textbf{72}(20), 205208 (2005)

\bibitem{daSilvaPinning}
C.J. da~Silva, J.P. Rino, L.~C\^andido, Phys. Rev. B \textbf{77}(16), 165407
  (2008)

\bibitem{damascenoPinning}
P.~Damasceno, C.~DaSilva, J.~Rino, L.~Cândido, J. Low Temp. Phys. \textbf{160},
  58 (2010)

\bibitem{Zimmerman}
N.M. Zimmerman, J.L. Cobb, A.F. Clark, Phys. Rev. B \textbf{56}(12), 7675
  (1997)

\bibitem{reesSCPT}
D.G. Rees, et~al., Appl. Phys. Lett. \textbf{93}(17), 173508 (2008)

\bibitem{mukharsky}
E.~Rousseau, et~al., Phys. Rev. B \textbf{79}(4), 045406 (2009)

\bibitem{Reesprl}
D.G. Rees, et~al., Phys. Rev. Lett. \textbf{106}(2), 026803 (2011)

\bibitem{Iye}
Y.~Iye, J. Low Temp. Phys. \textbf{40}(5-6), 441 (1980)

\bibitem{Flex}
The FEM modeling was performed using FlexPDE software, PDE Solutions Inc.

\bibitem{LeidererThinFilm}
J.~Angrik, A.~Faustein, J.~Klier, P.~Leiderer, J. Low Temp. Phys. \textbf{137},
  335 (2004)

\bibitem{ourJLTP}
D.G. Rees, K.~Kono, J. Low Temp. Phys. \textbf{158}(1-2), 301 (2010)

\bibitem{Hoogendoorn2005}
S.P. Hoogendoorn, W.~Daamen, Transportation Science \textbf{39}(2), 147 (2005)

\bibitem{OriginalQPC}
B.J. van Wees, et~al., Phys. Rev. Lett. \textbf{60}(9), 848 (1988)

\bibitem{Araki2011}
M.~{Araki}, H.~{Hayakawa}, ArXiv e-prints (arXiv:1104.4854)  (2011)

\bibitem{saitoh}
M.~Saitoh, T.~Aoki, Journal of the Physical Society of Japan \textbf{44}(1), 71
  (1978)

\bibitem{bridges}
F.~Bridges, J.F. McGill, Phys. Rev. B \textbf{15}(3), 1324 (1977)

\bibitem{ShikinMonarkhaDimples}
Y.P. {Monarkha}, V.B. {Shikin}, Sov. Phys. JETP \textbf{41}, 710 (1975)

\end{thebibliography}

\end{document}